\documentclass[9pt,5p,twocolumn]{elsarticle}

\makeatletter
\def\ps@pprintTitle{
 \let\@oddhead\@empty
 \let\@evenhead\@empty
 \def\@oddfoot{\it \hfill\today}
 \let\@evenfoot\@oddfoot}
\makeatother

\usepackage{amsmath}
\usepackage{graphicx}
\usepackage{dcolumn}
\usepackage{amsfonts}

\begin{document}

\author[add]{Axel P\'erez-Obiol}
\author[add]{Taksu Cheon}
\address[add]{Laboratory of Physics, Kochi University of Technology, Tosa Yamada, Kochi 782-8502, Japan}

\title{One dimensional Hydrogen problem with general connection condition at the origin
and non-Rydberg spectra} 
\date{today}

\begin{abstract}

We consider the solution of the quantum Coulomb problem in one dimension with the most
general connection condition at the origin. The divergence of the derivative of the 
wave function at the origin
invalidates the standard current conservation approach. We explore two approaches, 
Wronskian self-adjoint extension method and cutoff regularization method, and establish
their mutual relations, thereby clarifying the physical contents of the connection
parameters. We show how to realize exotic non-Loudon connection conditions,
entailing the realization of non-Rydberg spectrum.

\end{abstract}

\maketitle

\section{Introduction}

The one dimensional Coulomb problem, as defined by the 1D Schr\"odinger equation with a Coulomb potential,
\begin{align}
\label{eq:coulomb}
-\frac{\hbar^2}{2m}\,\psi''(x)-\frac{e^2}{|x|}\,\psi(x)
=E\,\psi(x),
\end{align}
is a useful model to describe physical systems
such as the hydrogen atom under strong
magnetic fields \cite{elliot60,rossner84} and the spectra of electrons bound above the surface of
 a superfluid \cite{cole69,cole70,care72}.

In contrast to the three-dimensional analog, the divergence of the potential at $x=0$ implies
that square integrable solutions have a divergent derivative at the origin.
Using atomic units, assuming bound states, $E<0$,
and defining $\alpha=\frac{1}{\sqrt{-2E}}$,
these solutions can be written in terms of
Whittaker functions,
\begin{align}
\label{eq:gensolution}
\psi(x)=N\left[\sin(\Omega)\theta(-x)+\cos(\Omega)\theta(x)\right]W_{\alpha,\frac12}\left(\tfrac{2}{\alpha}|x|\right),
\end{align}
with $N$ a normalization constant,
and where $\Omega\in[-\frac{\pi}{2},\frac{\pi}{2}]$ parametrizes the weight on the left and right sides of the singularity.
The divergence of $\psi'(x)$ at the origin does not allow for a trivial determination of the spectrum,
fixed by the possible values of $\alpha$ and $\Omega$.
On the one hand, unlike the case of the 1D point interaction on free space~\cite{albeverio05,shigehara98},
it is not possible to fix it in terms of a linear relation
between the wave functions and derivatives.
On the other, there is not a unique, physically valid condition, since in principle
different finite physical potentials might have different zero-range limits.

On a physical basis, various arguments have been made in order to support different
regularizations. In \cite{loudon59} two cut-offs were proposed, establishing a set
of odd and even solutions, with a ground energy diverging with the vanishing cut-off.
Since then, many works have discussed the existence of this state and the possible parity
of the solutions \cite{palma06}. In more mathematical terms, the connection condition has been constrained by
self-adjointness in \cite{fischer95,tsutsui02,oliveira09}.
In \cite{tsutsui02}, a general self-adjoint connection condition 
for singular potentials was developed.
It was defined in terms of Wronskians, and the corresponding constraints for the 1D Coulomb 
problem were provided as an example.

In this work, we pick up the Wronskian condition developed in \cite{tsutsui02},
apply it to the 1D Coulomb problem,
 and characterize both the energy spectra and possible eigenfunctions through 
the parameters defining the family of self-adjoint conditions (Sec.~\ref{sec:cc}).
The solutions are next analyzed in terms of the wave function behavior
near the origin, and in terms of a set of regularized potentials (Sec.~\ref{sec:potentials}).
These potentials consist in the Coulomb interaction at $|x|>d$,  plus three Dirac deltas at $x=-d,0,d$.
Their magnitudes are a function of the cut-off distance $d$, and depending how they behave in the
limit $d\to0$, they converge to one or another point interaction.
This construction is general enough so that we can relate the self-adjoint connection
condition to the behavior of the potentials, such as soft or hard cores, symmetric or antisymmetric (Sec.~\ref{sec:relation}). 
Thus, the 1D Coulomb problem is both general and rigorously formulated,
and at the same time intuitively understood on physical grounds.
A regularization with a rectangular barrier or well of width $d$ and centered at the origin
is discussed in Sec.~\ref{sec:rect}.
We conclude in Sec.~\ref{sec:conclusions}.

\section{Wronskian self-adjoint connection condition}
\label{sec:cc}

The general condition satisfying self-adjointness, as defined in \cite{tsutsui02}, reads,
\begin{align}
\label{eq:ucc}
 (U-I)\Psi+i(U+I)\Psi'=0,
\end{align}
where $U$ is a unitary 2x2 matrix, $I$ the identity, and
\begin{align}
 \Psi=
\left(\begin{matrix}
W[\psi,\phi_1]_{+} \\
W[\psi,\phi_1]_{-}
\end{matrix}\right),~~~
 \Psi'=
\left(\begin{matrix}
W[\psi,\phi_2]_{+} \\
-W[\psi,\phi_2]_{-}
\end{matrix}\right),
\end{align}
with $W[~]$ the Wronskian operator evaluated at $x\to 0_\pm$,
$W[f(x),g(x)]_{\pm}=f(x)g'(x)-f'(x)g(x)|_{x\to 0_\pm}$.
$\psi$, $\phi_1$ and $\phi_2$ are eigenfunctions,
$\psi$ being the solution to be constrained, defined in Eq.~(\ref{eq:gensolution}), and $\phi_1$ and $\phi_2$
auxiliary functions.
These functions must satisfy $W[\phi_1,\phi_2]=1$ and may have an arbitrary
eigenvalue $\beta$.
For the Coulomb 1D problem they can be defined as
\begin{align}
\phi_1(x)=&\frac{\beta}{2}M_{\beta,\frac12}\left(\tfrac{2}{\beta}x\right)(\theta(x)-\theta(-x)),
\\
\phi_2(x)=&-\Gamma(1-\beta)W_{\beta,\frac12}\left(\tfrac{2}{\beta}x\right),
\end{align}
with $M$ and $\Gamma$ the Whittaker and gamma functions.
We use the standard decomposition $U=V^{-1}\,D\,V$,
with $D$ diagonal and $V\in SU(2)$, and parametrize both matrices as
\begin{align}
D=&\left(\begin{matrix}
e^{-i\theta_+}  &  0 \\
0  & e^{-i\theta_-}
\end{matrix}
\right),
\\V=&\left(\begin{matrix}
e^{i\lambda}\cos(\omega)  &  \sin(\omega) \\
-\sin(\omega)  & e^{-i\lambda}\cos(\omega)
\end{matrix}
\right),
\end{align}
where $\theta_\pm,\lambda\in[-\pi,\pi)$
and $\omega\in[0,\pi)$.
A non-trivial solution of Eq.~(\ref{eq:ucc}) implies,
\begin{align}
\lambda=&0,
\\
\label{eq:tfcweight}
\omega=&\Omega+(1\mp1)\frac{\pi}{4},
\\
\label{eq:tfcspectrum}
\tan\left(\frac{\theta_\pm}{2}\right)=&
-\frac{1}{\alpha}+2\log(\alpha)-2F(1-\alpha),
\nonumber\\&
+\frac{1}{\beta}-2\log(\beta)+2F(1-\beta),
\end{align}
with $F$ the digamma function.
The most general self-adjoint connection condition in the bound 1D coulomb
problem can then defined by
three parameters, $\omega$, $\theta_-$, and $\theta_+$.
$\omega$ fixes the weight of the 
wave function in each side of the singularity, and $\theta_-$ and $\theta_+$
fix the spectrum $\alpha$. 
In the following, the arbitrary value of $\beta$ is fixed such that
\begin{align}
&\frac{1}{\beta}-2\log(\beta)+2F(1-\beta)+4\gamma+2\log(2)=0,
\\
\label{eq:tfcspectrum2}
&\tan\left(\frac{\theta_\pm}{2}\right)=
-\frac{1}{\alpha}-2\log\left(\frac{2}{\alpha}\right)-2F(1-\alpha)-4\gamma,
\end{align}
with $\gamma$ the Euler's constant.

\begin{figure}[t]
\centering
\includegraphics[width=.48\textwidth]{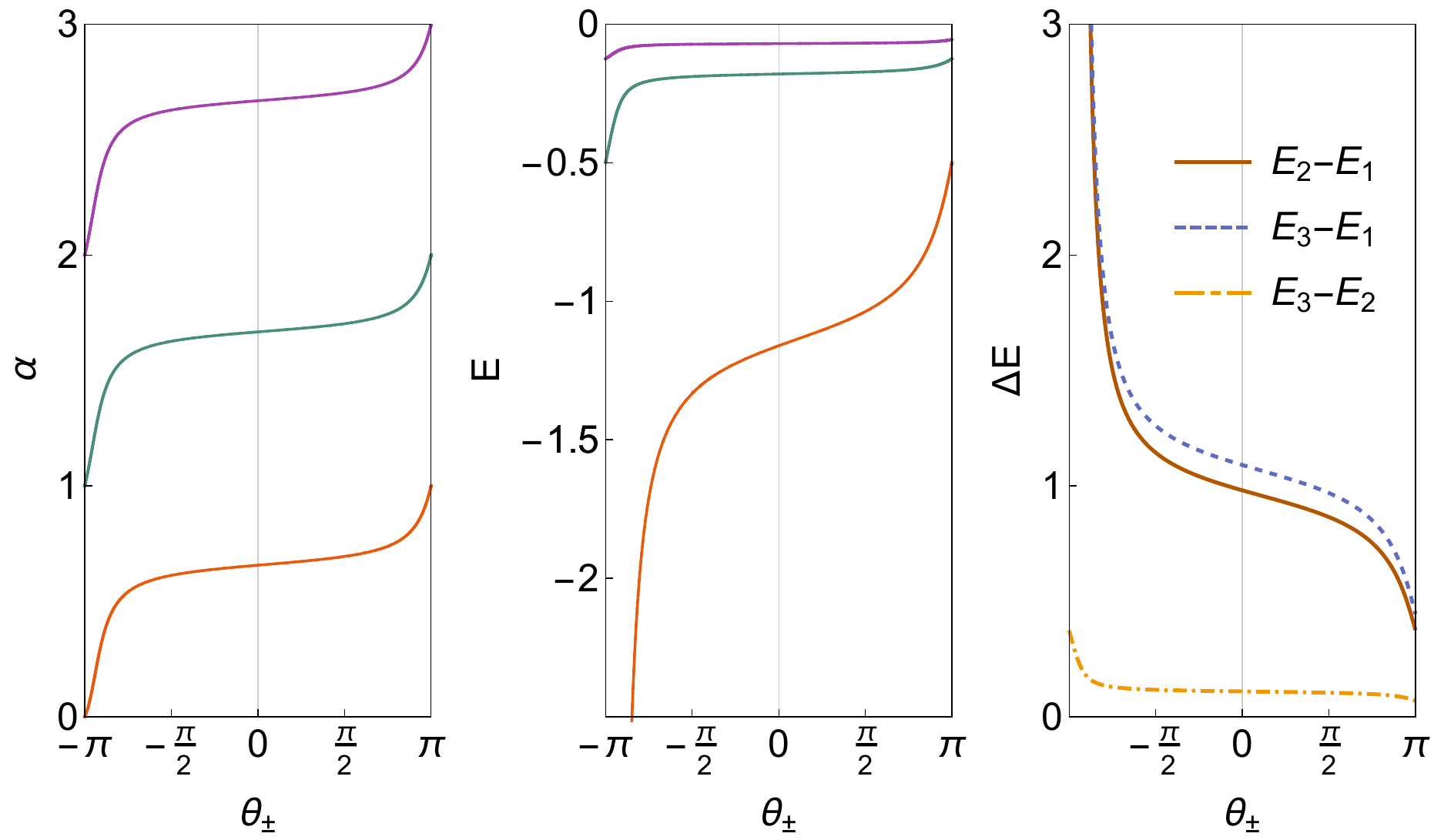}
\caption{(Color online). First three energy levels in terms of $\alpha=\frac{1}{\sqrt{-2E}}$ (left),
$E$ (middle), and their energy
gaps $\Delta E$ (right) as a function of point interaction parameters $\theta_\pm$.
The self-adjoint extension of the 1D Coulomb potential allows for two
independent spectra, each one parametrized by $\theta_-$ and $\theta_+$,
both relations $E(\theta_+)$ and $E(\theta_-)$ being the same.
The spectrum is continuous at $\theta_\pm=-\pi,\pi$, and at these values
it corresponds to the Rydberg series $\alpha=n$, $E=-\frac{1}{2\,n^2}$, with $n$ and integer.
}
\label{fig:spectrum}
\end{figure}

\begin{figure}[t]
\centering
\includegraphics[width=.48\textwidth]{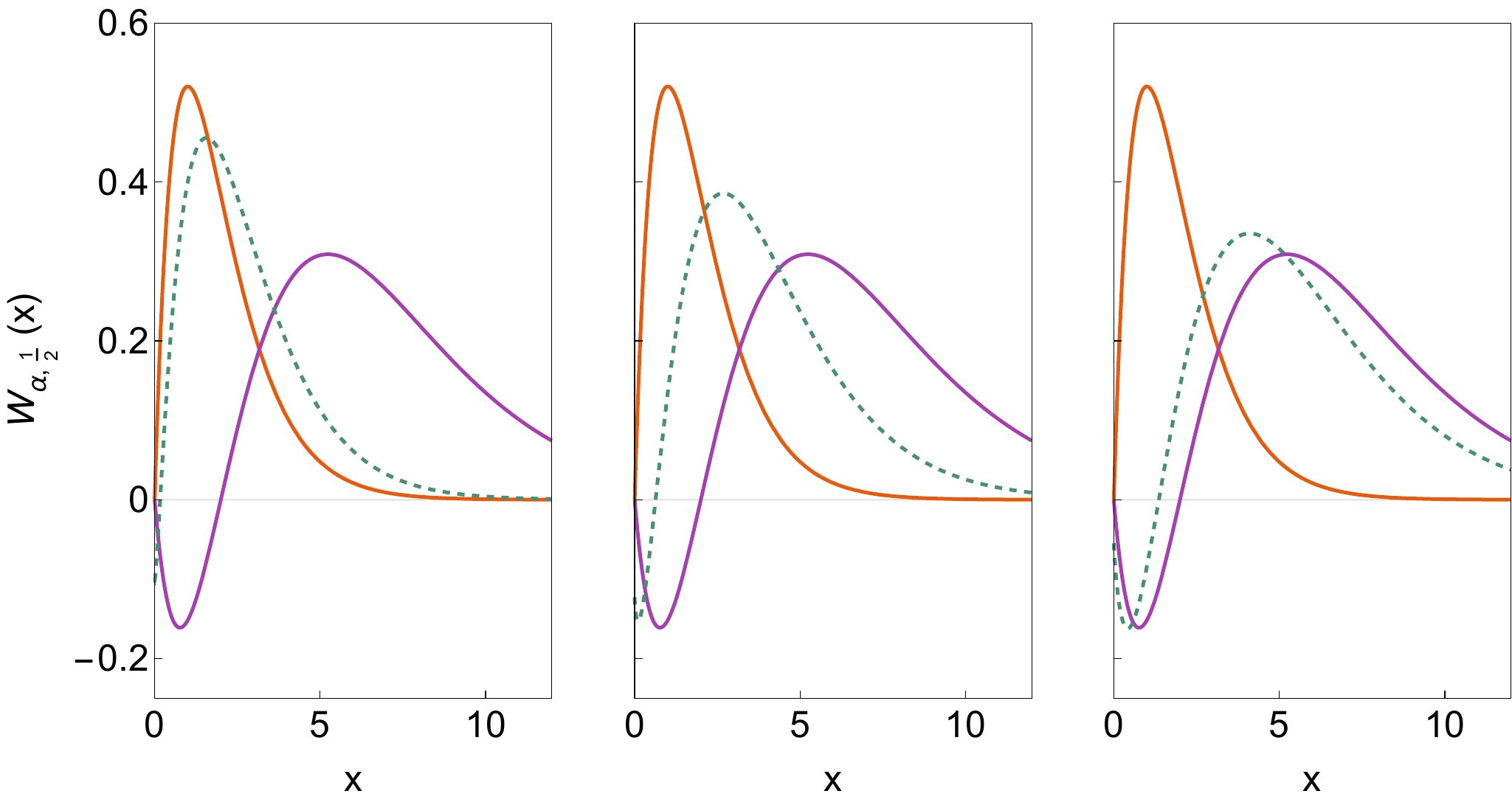}
\caption{(Color online). Normalized Whittaker function 
$N\,W_{\alpha,\frac12}\left(\tfrac{2}{\alpha}|x|\right)$
for $\alpha=1,2$ (solid red and solid purple in each panel),
and $\alpha=1.2,1.5,1.8$ (dashed green in the left, center,
and right panels, respectively).
}
\label{fig:plotw}
\end{figure}

Two spectra are possible depending on the values
of $\theta_\pm$, the relations $\alpha(\theta_{-})$ and $\alpha(\theta_+)$ being the same,
and $|\Omega(\theta_-)-\Omega(\theta_+)|=\frac{\pi}{2}$.
These spectra are periodic and continuous at $\theta_\pm=-\pi$ and $\theta_\pm=\pi$,
both points being degenerate and corresponding to Rydberg-type levels.
The ground state at $\theta_\pm=-\pi$ diverges, since $\alpha\to0$,
and is an exception to this degeneracy.
The lower part of the energy spectrum, in the form of $\alpha(\theta_\pm)$ and $E(\theta_\pm)$,
and the corresponding energy gaps, $\Delta E(\theta_\pm)$, are plotted in Fig.~\ref{fig:spectrum}.
The energy levels monotonously increase from $\alpha=n$ at $\theta_\pm=-\pi$,
to $\alpha=n+1$, at $\theta_\pm=\pi$, with $n\geq0$ integer.
Wave functions gradually become more peaked as $\theta_\pm$ decreases from
$\pi$ to $-\pi$, the ground state turning into a sharp singly or doubly peaked function
as $\theta_\pm\to-\pi$, and the rest converging back to the eigenfunctions at $\theta_\pm=\pi$.
The variation of $W_{\alpha,\frac12}\left(\tfrac{2}{\alpha}|x|\right)$
from $\alpha=1$ to $\alpha=2$ is illustrated in Fig.~\ref{fig:plotw}.

\section{Finite cut-offs}
\label{sec:potentials}

The 1D Coulomb potential can be regularized by cutting it off at
$x\in[-d,d]$, and then replacing the potential at this interval
by a finite one.
The connection condition at the origin is then replaced by boundary conditions at $x=\pm d$.
Before specifying any form of regularization, we analyze
how the spectrum depends on the behavior of the wave function at $x=\pm d$.
The weight $\Omega$ and energy spectrum $\alpha$ are independently related to the quotients,
\begin{align}
\label{eq:defr}
R(\Omega)\equiv&\frac{\psi(-d)}{\psi(d)}
=\tan(\Omega),
\\
Q(d,\alpha)\equiv&\frac{\psi'(d)}{\psi(d)}
\nonumber
=\frac{1}{\alpha}-\frac{\alpha}{d}-\frac{1}{d}\frac{W_{1+\alpha,\frac12}(\tfrac{2 d}{\alpha})}{W_{1+\alpha,\frac12}(\tfrac{2 d}{\alpha})}
\\=&
\nonumber
-\frac{1}{\alpha}-2\log\left(\frac{2d}{\alpha}\right)-2F(1-\alpha)-4\gamma
\\&
\label{eq:defq}
+{\mathcal O}\left(d\,(\log (d))^2\right).
\end{align}
In the limit $d\to0$, 
$Q$ and $\tan\left(\frac{\theta_\pm}{2}\right)$ have the same dependence on $\alpha$,
their difference being $2\log(d)$.
In general, a regularization allows for two sets of solutions,
as for example odd and even, entailing two possible $Q$ and $R$.
We label each set of solutions with subindices $-$ and $+$, and associate each one
to $\theta_-$ and $\theta_+$,
\begin{align}
\label{eq:rom}
R_\pm(\Omega)=&\tan\left(\omega+(1\mp1)\frac{\pi}{4}\right),
\\
\label{eq:qtan}
Q_\pm(d,\alpha)=&\tan\left(\frac{\theta_\pm}{2}\right)-2\log(d).
\end{align}
The connection parameters thus fix the behavior of the wave function at the origin.
$\theta_-$ and $\theta_+$ determine the value of $\lim_{d\to0}Q(d,\alpha)$ for each set of solutions,
while $\omega$ fixes the relative weights on each boundary $x=\pm d$, as $d\to0$.
Note that, in this limit, only in the case in which $Q(d,\alpha)$ diverges logarithmically,
$Q\sim -2\log(d)$, the spectrum will correspond to $\theta_\pm\neq\pm\pi$,
and therefore to a non-Rydberg spectrum.

We regularize Eq.~(\ref{eq:coulomb}) by replacing the Coulomb potential
at $x\in[-d,d]$ with three Dirac deltas at $x=-d,0,d$ of magnitudes $u_1$, $v$, $u_2$,
respectively,
and zero at $x\in(-d,0)$ and $x\in(0,d)$.
The free solutions at $x\in(-d,0)$ and $x\in(0,d)$ are a linear combination of exponentials $e^{\pm \frac{x}{\alpha}}$,
and assuming $d\ll\alpha$, they can be approximated to straight lines.
The values of $u_1$, $v$, and $u_2$ then fix $R$ and $Q$ to
\begin{align}
R_\pm(\Omega)=&
2\,d\,(1+d\,v)(u_2-u_1)
\nonumber\\&\pm\sqrt{1+(2\,d\,(1+d\,v)(u_2-u_1))^2},
\label{eq:tandel}
\\
Q_\pm(d,\alpha)=&
\frac{1}{2\,d}+u_1+u_2
\nonumber\\&
+\frac{d\,v\mp\sqrt{1+(2\,d\,(1+d\,v)(u_2-u_1))^2}}{2\,d\,(1+d\,v)}.
\label{eq:qdel}
\end{align}

For two exterior deltas of equal strength, $u_1=u_2=u$,
the above relations read,
\begin{align}
R_\pm(\Omega)=&\pm 1,
\\
Q_\pm(d,\alpha)=&2\,u+\frac{1}{d+\frac{1}{2\,v}(1\pm1)}.
\end{align}
In this case, the wave functions have well defined parity,
with $R_-(\Omega)=-1$ and $R_+(\Omega)=1$ corresponding to odd and even solutions.

\section{Relation among connection parameters and regularized potentials}
\label{sec:relation}

The behavior of the regularized potentials as $d\to0$, determining $R(\Omega)=\tan(\Omega)$
and $Q(d,\alpha)$, can be related to the connection parameters
$\theta_\pm$ and $\omega$ through Eqs.~(\ref{eq:rom}) and~(\ref{eq:qtan}).
For the case of three Dirac deltas, $R$ and $Q$ are given by
by Eqs.~(\ref{eq:tandel}) and~(\ref{eq:qdel}). Solving for $u_1$, $v$, and $u_2$, we find
\begin{align}
u_1=&
-\frac{1}{2\,d}-\log(d)
\nonumber\\&
+\frac14\left(1+\sin(2\,\omega)+\cos(2\,\omega)\right)\tan\left(\frac{\theta_{-}}{2}\right)
\nonumber\\&
+\frac14\left(1-\sin(2\,\omega)-\cos(2\,\omega)\right)\tan\left(\frac{\theta_{+}}{2}\right),
\label{eq:u1}
\\
v=&
-\frac{1}{d}
+\frac{1}{d^2}\,
\frac{\csc(2\,\omega)}{\tan\left(\frac{\theta_{-}}{2}\right)-\tan\left(\frac{\theta_{+}}{2}\right)},
\label{eq:v}
\\
u_2=&
-\frac{1}{2\,d}-\log(d)
\nonumber\\&
+\frac14\left(1+\sin(2\,\omega)-\cos(2\,\omega)\right)\tan\left(\frac{\theta_{-}}{2}\right)
\nonumber\\&
+\frac14\left(1-\sin(2\,\omega)+\cos(2\,\omega)\right)\tan\left(\frac{\theta_{+}}{2}\right).
\label{eq:u2}
\end{align}
Any spectrum of the self-adjoint extension of the 1D Coulomb problem,
defined by the three parameters $\theta_-$, $\theta_+$, and $\omega$,
can therefore be reproduced by two attractive
deltas at $x=\pm d$, and a middle one, repulsive or attractive, at $x=0$,
and vanishing cut-off distance $d$.
This relation can be checked by choosing a set of values for $\theta_\pm$ and $\omega$,
calculating the corresponding three deltas according to Eqs.~(\ref{eq:u1}),~(\ref{eq:v}), and~(\ref{eq:u2}),
and solving for the spectra $\alpha$ and $\Omega$ in the corresponding regularized 1D Coulomb problem.
The energy eigenvalues are found to converge to the ones given by the point interaction for $d\sim 10^{-4}$.
Note that, at this cut-off distance, the error in Eq.~(\ref{eq:defq}),
${\mathcal O}\left(d\,(\log (d))^2\right)$, is of the order of $1\%$.

\begin{figure}[t]
\centering
\includegraphics[width=.45\textwidth]{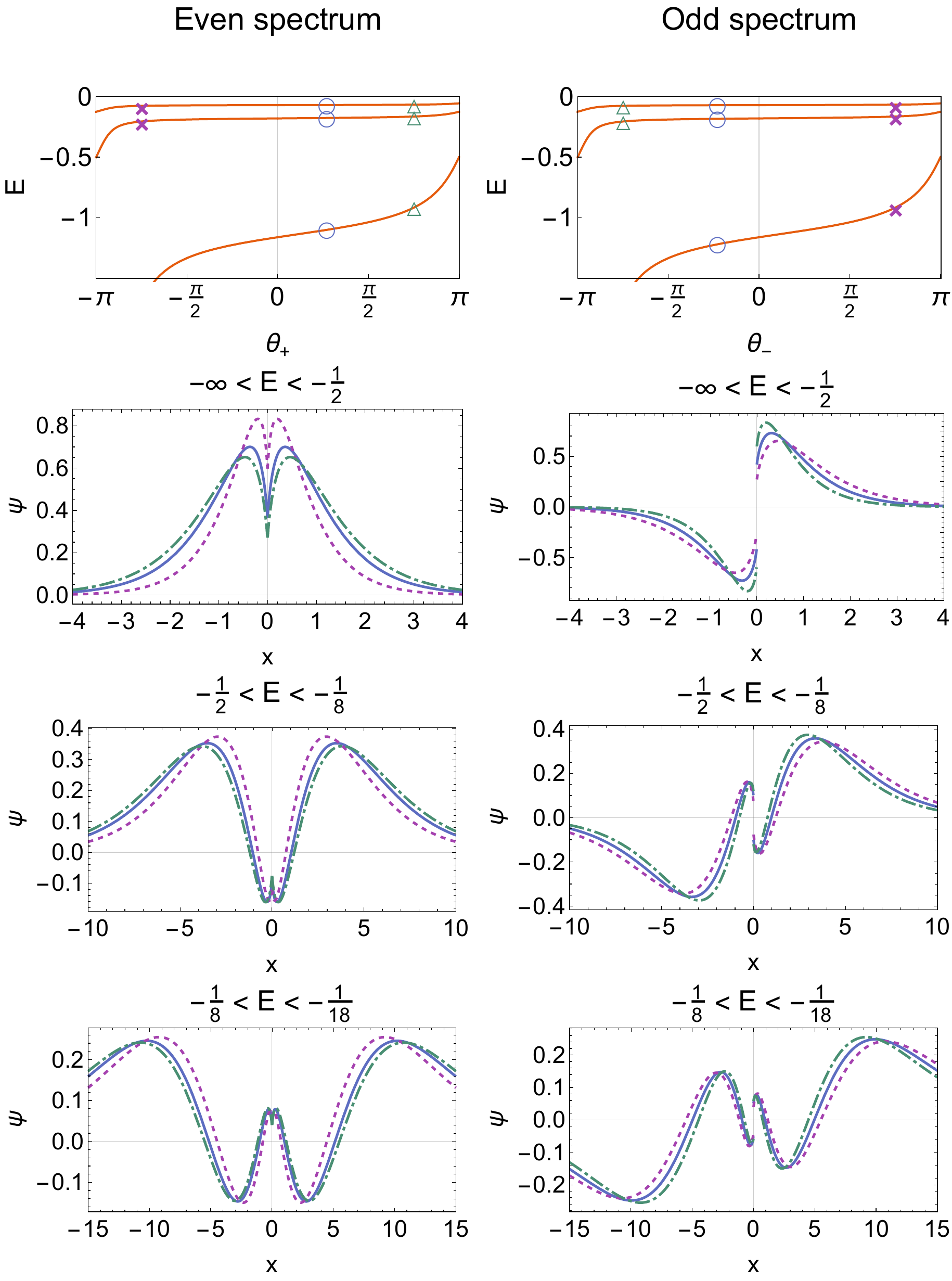}
\caption{
(Color online).
Top two plots: lower part of energy spectra (red color) $E(\theta_+)$ and $E(\theta_-)$
as a function of the point interaction parameters $\theta_+$ and $\theta_-$.
Plotted on top are three sets of energy values obtained through three different regularizations,
each consisting in the Coulomb potential at $|x|>d$,
and three Dirac deltas at $x=-d,0,d$, with $d=10^{-4}$. The deltas magnitudes
are given by Eqs.~(\ref{eq:u1}),~(\ref{eq:v}) and~(\ref{eq:u2}),
with $\theta_\pm=\pm\frac{\pi}{4}$ (circles),
$\theta_\pm=\pm\frac{3\pi}{4}$ (triangles),
$\theta_\pm=\mp\frac{3\pi}{4}$, (crosses). 
Triangle and cross markers corresponding to the ground state for the even
and odd spectra, respectively, are not shown due to their values being very low.
All have $\omega=\frac{\pi}{4}$,
which implies outer deltas of equal strength.
Bottom six plots: eigenfunctions corresponding to the same three eigenvalues and
for the three same regularizations,
solid lines corresponding to circles, dashed lines to  crosses, 
and dot-dashed lines to triangles.
They are organized according to the energy levels, $\frac{-1}{2n^2}<E<\frac{-1}{2(n+1)^2}$,
with $n$ integer.
}
\label{fig:spectradel}
\end{figure}

We illustrate how the energy spectrum and wave functions for the ground and first excited states
vary for three different symmetric regularizations (tending to three different
point interactions) for $d=10^{-4}$ in Fig.~\ref{fig:spectradel}.
All these regularizations correspond to $\omega=\frac{\pi}{4}$.
To obtain connection conditions with $\omega\neq\frac{\pi}{4}$,
the regularizing potentials must be asymmetric,
$u_2-u_1\propto\frac12\cos(2\,\omega)\neq0$.
This difference is maximal for $\omega=0,\frac{\pi}{2}$,
in which cases the wave functions domain reduces to the half line.

\section{Other regularizations}
\label{sec:rect}

The regularization of the 1D Coulomb potential with three Dirac deltas
has been chosen due to its simplicity, and such that is general
enough to map to the whole family of self-adjoint point interactions defined
in Sec.~\ref{sec:cc}.

Other regularizations are indeed possible. One example is the one first proposed in~\cite{loudon59},
consisting in a well of depth $V=-\frac{1}{d}$, width $2d$, and centered at the origin.
In this case the spectrum approaches $\alpha=n$ from above as $d\to0$,
with integer $n>0$ for even wave functions and $n\ge1$ for odd ones.
The energies are therefore lower than the corresponding Rydberg series.
In the limit $d\to0$, this regularization matches the point interaction with $\omega=\frac{\pi}{4}$,
$\theta_\pm=-\pi$. 
Another regularization studied in~\cite{loudon59}
is the Coulomb potential shifted from $x=-d,d$ towards $x=0$.
This implies odd wave functions with $0=\psi(d)\propto Q+2+\frac{1}{d}$, and even ones with
$0=\psi'(d)\propto Q$, therefore also converging to
$\omega=\frac{\pi}{4}$ and $\theta_\pm=-\pi$.
Both of these potentials diverge as $\frac{1}{d}$ at the cut-off distance,
and behave as Coulomb for $x\gg d$.
This is what is expected when integrating out the transversal degrees of freedom
in the dynamics of a hydrogen atom under strong magnetic fields~\cite{loudon16}.

A rectangular potential of general depth or height $V(d)$ may be considered.
The boundary conditions at $x=\pm d$,
resulting from matching the solutions and their derivatives inside
and outside the well or barrier, read,
\begin{align}
R(\Omega)&=-1,~~&Q(d,\alpha)=k\coth(k\,d);
\label{eq:vodd}
\\
\label{eq:veven}
R(\Omega)&=1,~~&Q(d,\alpha)=k\tanh(k\,d);
\end{align}
corresponding to odd and even solutions, respectively,
and where $k=\sqrt{2(V(d)-E)}$.
The spectrum is then found by 
replacing in the above equations the expressions for $R(\Omega)$ and $Q(d,\alpha)$,
defined in Eqs.~(\ref{eq:defr}) and~(\ref{eq:defq}), and solving for $\Omega$ and $\alpha$.
The particular case studied in~\cite{loudon59} corresponds to $V_1(d)=-\frac1d$.

\begin{figure}[t]
\centering
\includegraphics[width=.46\textwidth]{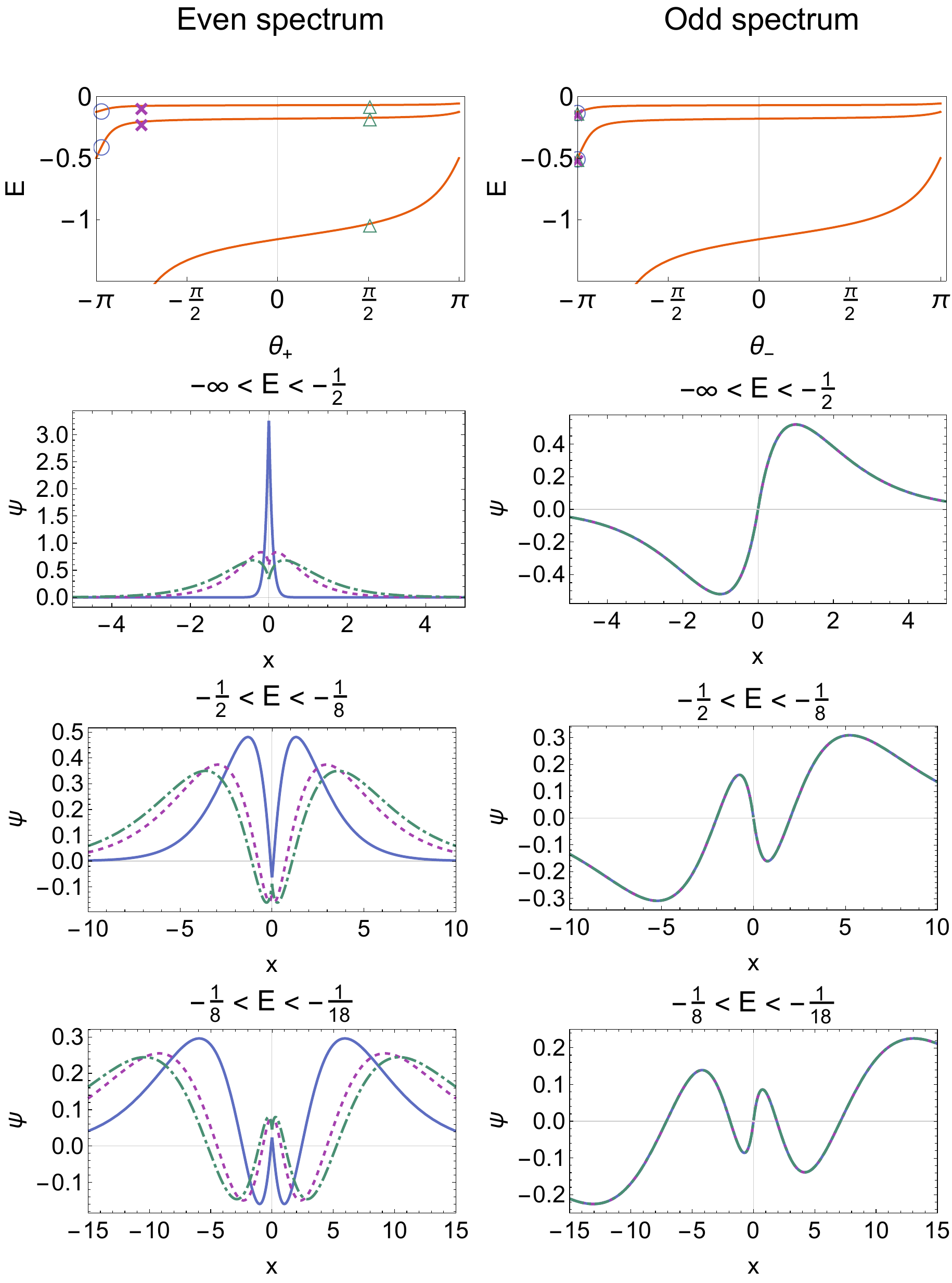}
\caption{(Color online).
Eigenvalues (top two panels) and eigenfunctions (bottom six panels)
for the ground state and first excited states
for three different rectangular potential regularizations.
They consist in the Coulomb potential at $|x|>d$, and a constant one
at $x\in[-d,d]$, of magnitude
$V_1(d)=-\frac{1}{d}$ (circles and solid lines),
and $V_2(d)=\frac{1}{d}\left(-\log(d)+\frac12\tan\left(\frac{\theta_+}{2}\right)\right)$,
with $\theta_+=-\frac{3\pi}{4}$ (crosses and dashed lines),
and $\theta_+=\frac{\pi}{2}$ (triangles and dot-dashed lines).
The cut-off distance has been set to $d=10^{-4}$, for which the regularizations
converge to the point interactions with $\omega=\frac{\pi}{4}$,
$\theta_-=-\pi$, and $\theta_+=-\pi,-\frac{3\pi}{4},\frac{\pi}{2}$.
Only the even spectrum has solutions for the lowest energy level, 
at $-\infty<E<-\frac{1}{2}$.
For $\theta_+=-\pi,-\frac{3\pi}{4}$ (circles and crosses),
the energy values are off the scale in the top-left plot.
}
\label{fig:spectrarect}
\end{figure}

Regularizing the potential through a well or barrier of the type
$V_2(d)=\frac{1}{d}\left(-\log(d)+\frac12\tan\left(\frac{\theta}{2}\right)\right)$,
with $\theta\in[-\pi,\pi)$,
the boundary conditions~(\ref{eq:vodd}) and~(\ref{eq:veven}) become,
\begin{align}
R(\Omega)=-1,~~Q(d,\alpha)\simeq & \frac{1}{d}-\frac23\log(d)+\frac13\tan\left(\frac{\theta}{2}\right);
\label{eq:vodd2}
\\
\label{eq:veven2}
R(\Omega)=1, ~~Q(d,\alpha)\simeq &-2\log(d)+\tan\left(\frac{\theta}{2}\right).
\end{align}
Using the definitions in Eqs.~(\ref{eq:rom}) and~(\ref{eq:qtan})
in the above equations, and taking the limit $d\to0$,
the above regularization maps to the subfamily of connection conditions given by
$\omega=\frac{\pi}{4}$, $\theta_-=-\pi$, and $\theta_+=\theta$.
The spectrum of odd wave functions converges then to $\alpha=n$,
with $n\ge1$.
The even spectrum gradually varies from $\theta_+=-\pi$, for either wells diverging
as $V=-\frac{1}{d}$ or slower, and soft barriers,
through $\theta_+=0$, for a barrier with height diverging as $V=-\frac{\log(d)}{d}$,
and to $\theta_+=\pi$ for hard core barriers.
Three samples of such spectra are illustrated in Fig.~\ref{fig:spectrarect}.

\section{Conclusions}
\label{sec:conclusions}

\noindent

We have analyzed the spectrum of the 1D Coulomb problem with a general
self-adjoint connection condition. This connection condition is defined by
three parameters, $\omega$, $\theta_-$, and $\theta_+$.
$\omega$ is directly related to the weight of the wave function on the left and right sides
of the point interaction, while $\theta_-$ and $\theta_+$ independently fix the energy spectrum
and shape of the wave functions, each parameter corresponding to a different set of solutions.
Based only on self-adjointness of the point interaction, for
symmetric boundary conditions, $\omega=\frac{\pi}{4}$,
odd and even solutions,
and the limit $E\to-\infty$ are allowed. Moreover, there is a continuum of possible
spectra, which, in general, do not have a well defined parity, differ from the Rydberg-type series,
and are not degenerate.

In order to provide physical meaning to the self-adjoint family of point interactions,
we have related  $\omega$ and $\theta_\pm$ to a set of regularized potentials.
They consist in a Coulomb potential which is cut-off at $x\in[-d,d]$, and replaced in this interval
by three Dirac deltas at $x=-d,0,d$.
These potentials are characterized by the three strengths of the deltas
and the cut-off distance $d$.
Depending on how the delta magnitudes behave as the cut-off distance vanishes,
the regularized potential tends to one or another set of connection parameters.

The case of single well or barrier with varying depth or height has also been analyzed,
and related to connection conditions with $\omega=\frac{\pi}{4}$, $\theta_-=-\pi$ and free $\theta_+$.
In general, regularized potentials which imply boundary conditions with
$Q=\frac{\psi'(d)}{\psi(d)}=\tan\left(\frac{\theta_\pm}{2}\right)-2\log(d)$, map
to point interactions with $\theta_\pm$.

\section*{Acknowledgments}
We acknowledge the support by
KUT presidential grant at Research Institute, Kochi University of Technology.

\clearpage
\end{document}